\documentclass[prl,reprint,aps,showpacs,amsmath,amssymb,showkeys,nofootinbib,floatfix]{revtex4-1}
\usepackage[english]{babel}
\usepackage{txfonts}
\usepackage{graphicx}

\newcommand{\gafd}{Geophys.\ Astrophys.\ Fluid\ Dyn.}  
\newcommand{\apjl}{Astrophys. J. Lett.}  
\newcommand{\pof}{Phys. Fluids}  
\newcommand{\pop}{Phys. Plasmas}  
  
\newcommand{\gji}{Geophys. J. Int.}  
\newcommand{\jfm}{J. Fluid Mech.}  
\newcommand{\njp}{New J. Phys.}  
\newcommand{\ssr}{Space Sci. Rev.}  
\newcommand{\prf}{Phys. Rev. Fluids}  

\renewcommand{\vec}[1]{\mbox{\boldmath$#1$}}

\begin{document}

\preprint{APS/123-QED}

\title[Dynamo action in a precessing cylinder]{Nonlinear 
  large scale flow in a precessing cylinder and its ability to
  drive dynamo action} 

\author{Andr\'e Giesecke}
\email{a.giesecke@hzdr.de}
\author{Tobias Vogt}
\author{Thomas Gundrum}
\author{Frank Stefani}
\affiliation{
Helmholtz-Zentrum Dresden-Rossendorf\\
Bautzner Landstrasse 400, D-01328 Dresden, Germany
}

\date{\today}

\begin{abstract}
{
We have conducted experimental measurements and numerical simulations
of a precession driven flow in a cylindrical cavity.  The study is
dedicated to the precession dynamo experiment currently under
construction at Helmholtz-Zentrum Dresden-Rossendorf (HZDR) and aims
at the evaluation of the hydrodynamic flow with respect to its ability
to drive a dynamo. We focus on the strongly nonlinear regime in which
the flow is essentially composed of the directly forced primary Kelvin
mode and higher modes in terms of standing inertial waves arising from
nonlinear self-interactions.  We obtain an excellent agreement
between experiment and simulation with regard to both, flow amplitudes
and flow geometry. A peculiarity is the resonance-like emergence of an
axisymmetric mode that represents a double roll structure in the
meridional plane.  Kinematic simulations of the magnetic field
evolution induced by the time-averaged flow yield dynamo action at
critical magnetic Reynolds numbers around ${\rm{Rm}}^{\rm{c}}\approx
430$ which is well within the range of the planned liquid sodium
experiment.
}
\end{abstract}


\maketitle

Magnetic fields of celestial bodies like planets, moons or asteroids
are ubiquitous in the solar system with a wide diversity of
manifestations \cite{wicht2010}.  While it is undisputed that these
fields are generated by conversion of mechanical energy from the flow
of an electrically conductive fluid, there are various possibilities
to drive the underlying fluid motions.  Usually, it is assumed that
the flow in liquid planetary cores is driven by thermo-compositional
convection \cite{braginsky1995}, yet alternative approaches invoke
mechanical stirring by libration \cite{wu2013}, tidal forcing
\cite{cebron2014} or precession \cite{tilgner2005}.  In particular
precession has been repeatedly proposed as source for dynamo action of
the ancient lunar magnetic field \cite{dwyer2011,*noir2013,*weiss2014}
or the geodynamo \cite{malkus1968,*vanyo1991}. Indeed, simulations
and experiments revealed that precession may excite vigorous flows
\cite{gans1970,*manasseh1992,*manasseh1996,*noir2003,*lagrange2011,*goto2014,*lin2014,*lin2015,*marques2015,*giesecke2015b,*lopez2016,*mouhali2012}
which are supposed to drive a dynamo \cite{lebars2015}.  Precessional
forcing has become of great interest from the experimental point of
view, because it represents a natural mechanism which allows an
efficient driving of conducting fluid flows on the laboratory scale
without making use of propellers or pumps
\cite{leorat2003,*leorat2006,*stefani2008}.  At HZDR a precession
dynamo experiment is under development \cite{stefani2012} which will
provide a flow of liquid sodium in a cylindrical cavity with a
magnetic Reynolds number of up to
${\rm{Rm}}\!=\!\varOmega_{\rm{c}}R^2/\eta\!\approx\!700$ (defined with
the achievable angular velocity of the cylinder
$\varOmega_{\rm{c}}\!=\!63\mbox{ s}^{-1}$, the radius $R\!=\!1\mbox{
m}$, and the magnetic diffusivity for liquid sodium
$\eta\!=\!0.09\,{\rm{m}}^2/{\rm{s}}$).  The project is further
motivated by previous precession experiments
conducted by \citet{gans1971} who achieved an amplification of an
applied magnetic field by a factor of 3 with a device smaller by a
factor of 8 and by numerical studies yielding precession driven
dynamos in different geometries with a critical magnetic Reynolds
number of $O(10^3)$
\cite{tilgner2005,wu2009,*hullermann2011,*lin2016}.  However, so far
numerical models of the planned experiment have not shown conclusively
that the achievable magnetic Reynolds number will be sufficient to
allow for dynamo action \cite{giesecke2015a,nore2011,*goepfert2016}.

In the present study we address preparatory simulations and
flow measurements at a water experiment that represents a down-scaled
model of the planned sodium dynamo.  The results provide flow patterns
and amplitudes in dependence on Reynolds number
${\rm{Re}}\!=\!\varOmega_{\rm{c}}R^2/\nu$ and on the relation of
precession frequency $\varOmega_{\rm{p}}$ to rotation frequency
$\varOmega_{\rm{c}}$, the Poincar{\'e} number
${\rm{Po}}\!=\!\varOmega_{\rm{p}}/{\varOmega_{\rm{c}}}$.  Finally, the
three-dimensional velocity fields from the simulations are used in
kinematic dynamo models in order to estimate parameter regimes that
will be appropriate for dynamo action.

\begin{figure*}[!t]
\includegraphics[width=2.05\columnwidth]{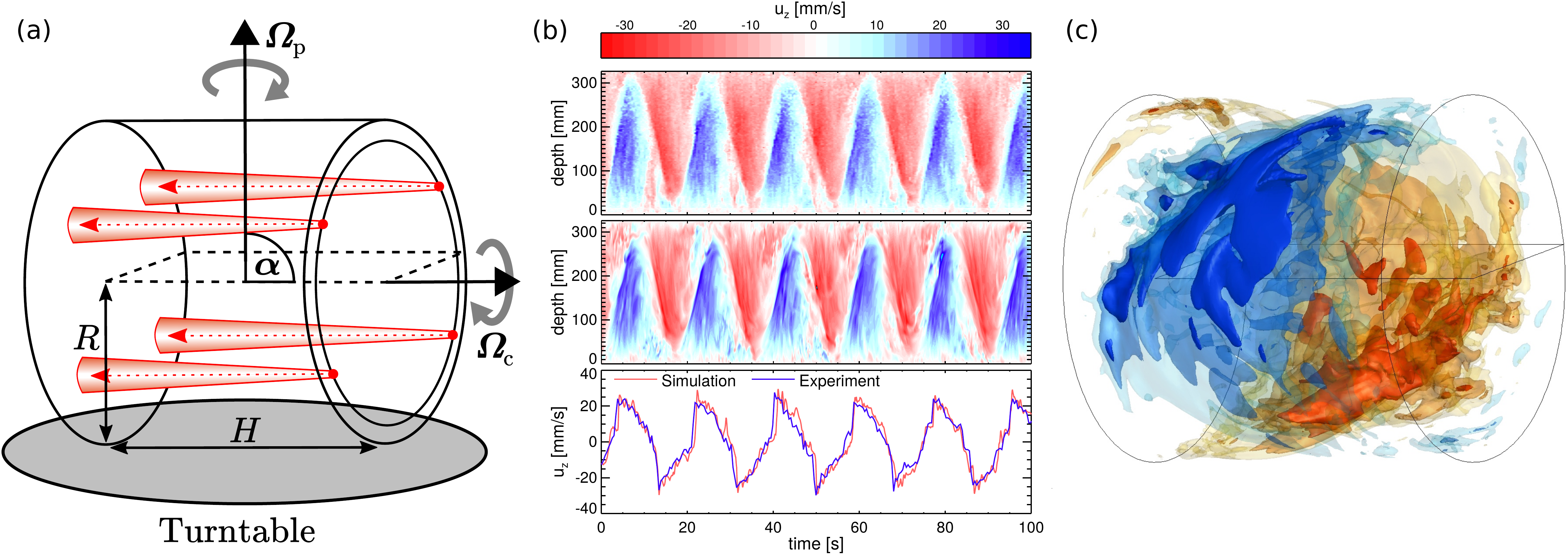}
\caption{
(a) Sketch of the experimental set-up. The red dots denote the
locations of the UDV probes in the water experiment and the arrows
illustrate the propagation of the ultrasound signal.
(b) Temporal evolution of the axial velocity $u_z$ at
$r\!=\!150\,{\rm{mm}}$ (top: UDV measurements, center: simulations,
bottom: comparison of simulations and experiments in the equatorial
plane).
(c) Isosurfaces showing a snapshot of $u_z$ from simulations at
${\rm{Re}}\!=\!10^4$ and ${\rm{Po}}\!=\!0.1$. Blue (red) colors
indicate flow in negative (positive) direction (see movie at
\cite{movie01} for temporal evolution of $u_z$).
\label{fig::sketch}} 
\end{figure*}

We conduct direct numerical simulations in the precession reference
frame using the code {\it{SEMTEX}} \cite{blackburn2004}. In this frame
the observer resides on the turntable following the rotation around
the precession axis, thereby watching at the spinning cylinder
(Fig.~\ref{fig::sketch}a). The flow is described by the Navier Stokes
equation including a time-independent term for the Coriolis force due
to precession \cite{tilgner1998}:
\begin{equation}
\frac{\partial}{\partial t} \vec{u}+\vec{u}\nabla\vec{u}
=-\nabla P-2\vec{\varOmega_{\rm{p}}}\times\vec{u}+\nu\nabla^2\vec{u}.
\label{eq::navier} 
\end{equation}
Here, $\vec{u}$ is the incompressible velocity field, $P$ the reduced
pressure, $\nu$ the viscosity, and $\vec{\varOmega}_{\rm{p}}$ the
angular velocity of the precessional motion.  The flow obeys no-slip
boundary conditions for the poloidal components, $u_r=u_z=0$, whereas
the azimuthal flow at the boundaries is prescribed by
$u_{\varphi}=r\varOmega_{\rm{c}}$.  Fluid velocities are measured
using {\it{Ultrasonic Doppler Velocimetry}} (UDV) which provides
instantaneous profiles of the velocity component in direction of an
ultrasonic beam \cite{takeda1986,vogt2014} oriented parallel to the
cylinder axis.  Four ultrasound transducers are fixed at one end cap
of the cylinder (Fig.~\ref{fig::sketch}a) and co-rotate with the
container thus providing measurements in the cylinder frame.  This
reference frame is well suited for flow characterization in terms of
eigenmodes of rotating flows which are the solutions of the linearized
inviscid version of Eq.~(\ref{eq::navier}). In a cylinder
these solutions are inertial waves, or Kelvin modes,
$\vec{U}_{\!j}(r,z,\varphi,t)=\tilde{\vec{u}}_{\!j}(r)e^{i(\omega_{\!j}t+m\varphi+kz)}$
\cite{kelvin1880,greenspan} labeled by $j$ that abbreviates a triple
index comprising the azimuthal wave number $m$, the axial wave number
$k$, and a radial wave number index $n$. The last index counts the
roots of the dispersion relation for an inertial wave
\begin{equation}
\omega_{\!j}\lambda_{\!j} J_{\!m-1}(\lambda_{\!j}) +
m\!\left(2\!-\!\omega_{\!j}\right)\!J_{\!m}(\lambda_{\!j})\!=\!0 
\mbox{ with }
\omega_{\!j}\!=\!
\displaystyle\frac{\pm 2}{\sqrt{1\!+\!\left(\frac{\lambda_{\!j}}{2 k\pi}\right)^2}}
\label{eq::dispersion}
\end{equation}
where $J_m$ denotes the Bessel function of order $m$, and
$\lambda_{j}$ plays the role of a radial wave number.

Precession causes a steady volume forcing with an odd symmetry with
respect to the equatorial plane.  Therefore the primary response of
the fluid is a flow with an azimuthal wave number $m=1$ and an odd
axial wave number $k$ that is stationary in the precession reference
frame. If the frequency of the corresponding eigenmode ($\omega_j$)
exactly matches the frequency of the forcing ($\varOmega_{\rm{c}}$),
the mode becomes resonant, and the linear inviscid approach for the
computation of the amplitude fails \cite{liao2012}.  The resonance
condition delicately depends on the aspect ratio, and the primary
forced mode with the simplest possible structure, i.e. $m=1$, $k=1$
and $n=1$ becomes resonant at $H/R=1.98982$ which is close to the
geometry envisaged for our planned experiment ($H/R=2$).  In the
present study the corresponding cylinder utilized in the water
experiment has radius $R=163\,{\rm{mm}}$ and height
$H=326\,{\rm{mm}}$, and the angle between rotation axis and precession
axis is fixed at $\alpha=90^{\circ}$.  Typical measurements of a
single UDV probe are shown in Fig.\ref{fig::sketch}b (top) in terms of
the axial velocity versus time and depth.  The alternation of the sign
of $u_z$ with the periodicity of $\varOmega_{\rm{c}}$ and the
asymmetry with respect to the equatorial plane illustrate the
dominance of the $m=1$ component superposed by higher azimuthal modes
(essentially $m=2$).  We find a very good agreement between
experiments and simulations (Fig.~\ref{fig::sketch}b, central and
bottom panel).  For sufficiently large ${\rm{Po}}$, the flow is
concentrated in the vicinity of the cylinder walls
(Fig.~\ref{fig::sketch}c) and can be decomposed into few large scale
modes. These modes represent standing inertial waves in the precession
reference frame, and time-dependent contributions only appear as weak
small-scale fluctuations (see movie in supplementals \cite{movie01}).

A quantitative analysis of the flow is done by decomposing axial
profiles of $u_z$ in $k$-modes $\propto \sin(\pi zk/H)$ which is the
characteristic $z$-dependence of the axial component of an inertial
wave in a cylinder with height $H$ \cite{meunier2008,liao2012}.  In a
second step we take the individual $k$-modes from this decomposition
and apply a 2D Fourier transformation in azimuthal direction and in
time which finally yields spectra that allow the identification of
individual modes labeled by $(m,k)$.
\begin{figure*}[t!]
\includegraphics[width=2.1\columnwidth]{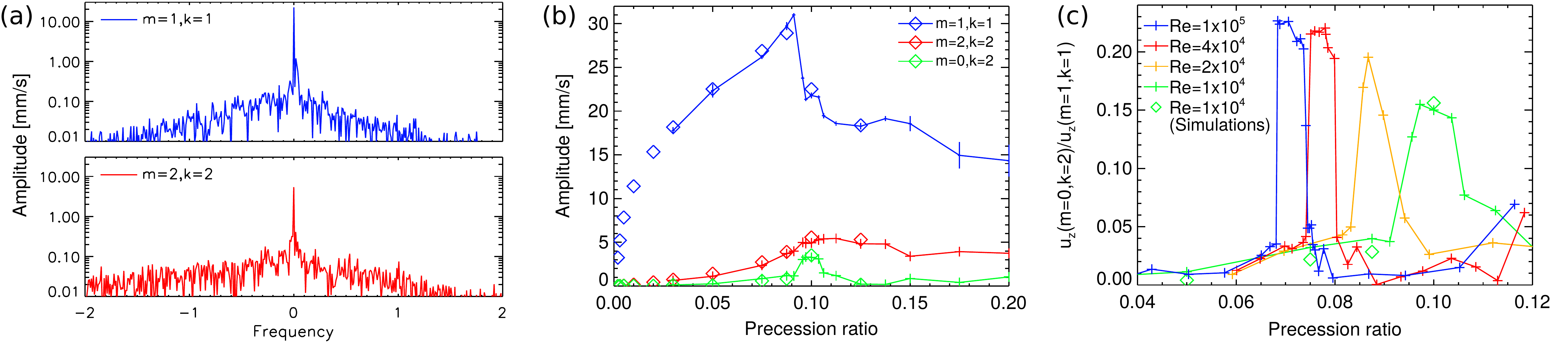}
\caption{
(a) Fourier spectra for the $(m,k)=(1,1)$ mode and for the
strongest secondary mode $(m,k)=(2,2)$ from simulations at
$r=150\mbox{ mm}$, ${\rm{Re}}=10^4$ and ${\rm{Po}}=0.1$.  (b)
Amplitude of the time-independent part of directly forced mode
$(m,k)=(1,1)$, its multiple $(m,k)=(2,2)$ and the non-geostrophic
axisymmetric mode $(m,k)=(0,2)$ (${\rm{Re}}=10^4$,
$r=150\mbox{ mm}$).  (c) Relative amplitude of the non-geostrophic
axisymmetric mode $(m,k)=(0,2)$ with respect to $(m,k)=(1,1)$.  
The solid curves in (b) and (c) denote results from the
water experiment, and the diamonds denote results from simulations.
\label{fig::amp}}
\end{figure*}
Typical spectra from simulations at ${\rm{Re}}=10^4$ and
${\rm{Po}}=0.1$ are shown in Fig.~\ref{fig::amp}a which represents the
signature of the primary forced mode $(m,k)=(1,1)$ and its first
multiple $(m,k)=(2,2)$ resulting from nonlinear self-interaction.  For
sufficiently strong precession the spectra of all $(m,k)$-modes
qualitatively look similar with one single peak at $\omega=0$ that
corresponds to a standing inertial wave in the precession reference
frame.  The amplitudes of individual modes, estimated from
spectral peaks at $\omega=0$, show that, independent of
${\rm{Po}}$, the flow is always dominated by the primary forced mode
$(m,k)=(1,1)$ (Fig.~\ref{fig::amp}b, blue curve).  A characteristic
feature is the concise maximum of the amplitude at
${\rm{Po}}^{\rm{c}}\approx 0.09$.  Immediately following this maximum
we find three phenomena that are intimately connected: a strong and
abrupt reduction of the amplitude of the directly forced flow with
$(m,k)=(1,1)$, a gradual increase of higher modes that originate from
nonlinear self-interaction according to $(m,k)\rightarrow (2m,2k)$
(Fig.~\ref{fig::amp}b, red curve) and a sudden appearance of a
non-geostrophic axisymmetric flow with $k$ even (Fig.~\ref{fig::amp}b,
green curve).
\begin{figure}[b!]
\includegraphics[width=0.9\columnwidth]{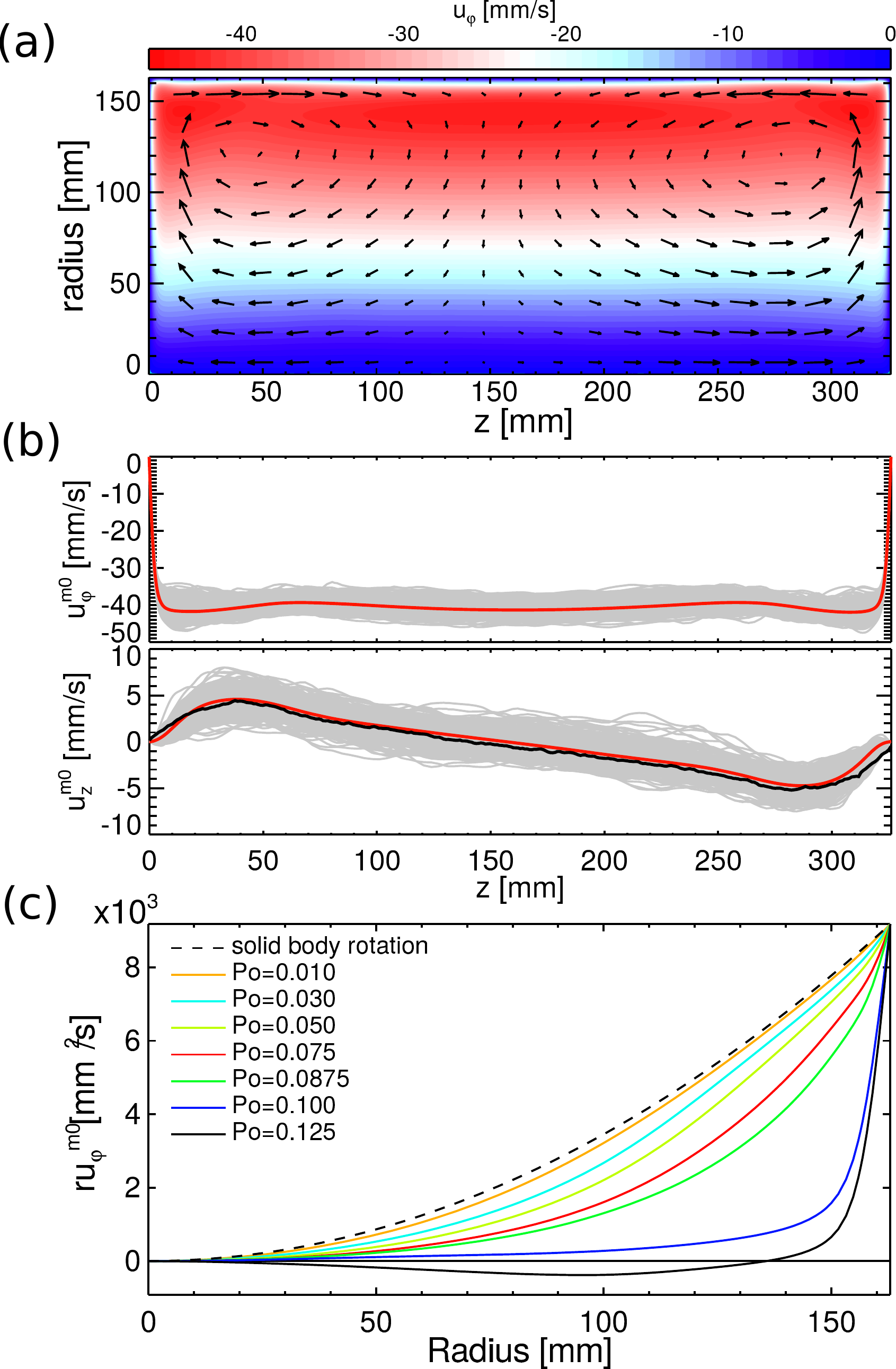}
\caption{
(a) Time-averaged axisymmetric velocity field at ${\rm{Re}}=10^4$ and
${\rm{Po}}=0.1$. Colors denote $u_{\varphi}$ (without solid body
rotation) and  arrows represent $u_r$ and $u_z$.
(b) Axial profile of $u_{\varphi}$ and $u_z$ at $r=150\,{\rm{mm}}$.
Grey curves represent temporal variations of instantaneous
profiles from simulations and red curves show the 
time average. The black curve in the bottom panel shows the
time-averaged profile obtained in the water experiment.
(c) Radial profiles of the time-averaged angular momentum including
solid body rotation from simulations at ${\rm{Re}}=10^4$.
\label{fig::m0struc}}
\end{figure}
The axisymmetric mode only exists with noteworthy amplitude within a
rather narrow band with a width $\Delta {\rm{Po}}\!\!\sim\!\!0.006$
(Fig.~\ref{fig::amp}c).  This axisymmetric mode is of interest with
regard to the dynamo problem because its geometric pattern corresponds
to a double roll structure (Fig.~\ref{fig::m0struc}a) similar to the
mean poloidal flow in the von-K{\'a}rm{\'a}n sodium dynamo in
which the flow was driven by two opposite counter-rotating impellers
\cite{monchaux2007}.  It is well known that this flow can drive a
dynamo at comparatively low ${\rm{Rm}}$ \cite{dudley1989,*ravelet2005}
when the relation between toroidal and poloidal components is of order
unity.  However, there are further contributions to the axisymmetric
flow in terms of a geostrophic azimuthal circulation
(Fig.~\ref{fig::m0struc}b) directed opposite to the solid body
rotation which worsen this relation in our model.

The experiments show that the basic flow properties remain
unchanged up to ${\rm{Re}}=10^5$ except the decrease of the critical
value ${\rm{Po}}^{\rm{c}}$ at which the previously discussed phenomena
emerge. The occurrence of the non-geostrophic axisymmetric resonance
is a robust feature which does not disappear when
increasing ${\rm{Re}}$ (Fig.~\ref{fig::amp}c).  This mode can be
excited by interacting inertial waves according to
$(m,k,\omega)\rightarrow(0,2k,0)$ \cite{meunier2008}.  However, this
is unlikely without the presence of singularities \cite{greenspan1969}
so that these interactions must happen within no-slip boundary layers
\cite{busse1968} or internal shear layers \cite{tilgner2007b}.  A more
descriptive explanation rests upon the modification of the basic
azimuthal circulation, which for sufficiently large ${\rm{Po}}$
compensates the bulk fluid's solid body rotation.  The azimuthal fluid
motion opposite to the cylinder rotation can even become so strong
that eventually the Rayleigh criterion for stability of rotating
fluids may be violated by developing a negative radial derivative of
the angular momentum, i.e., $\frac{d}{dr}(u_{\varphi}r) < 0$
(Fig.~\ref{fig::m0struc}c), immediately leading to the formation of
Taylor vortices.
Finally, the further increase of ${\rm{Po}}$ leads to the breakdown of the
large scale structures into smaller scales which, at ${\rm{Re}}\approx
O(10^6)$, corresponds to a transition into a fully turbulent flow without
significant large scale contributions \cite{herault2015}.

\begin{figure}[t!]
\includegraphics[width=1\columnwidth]{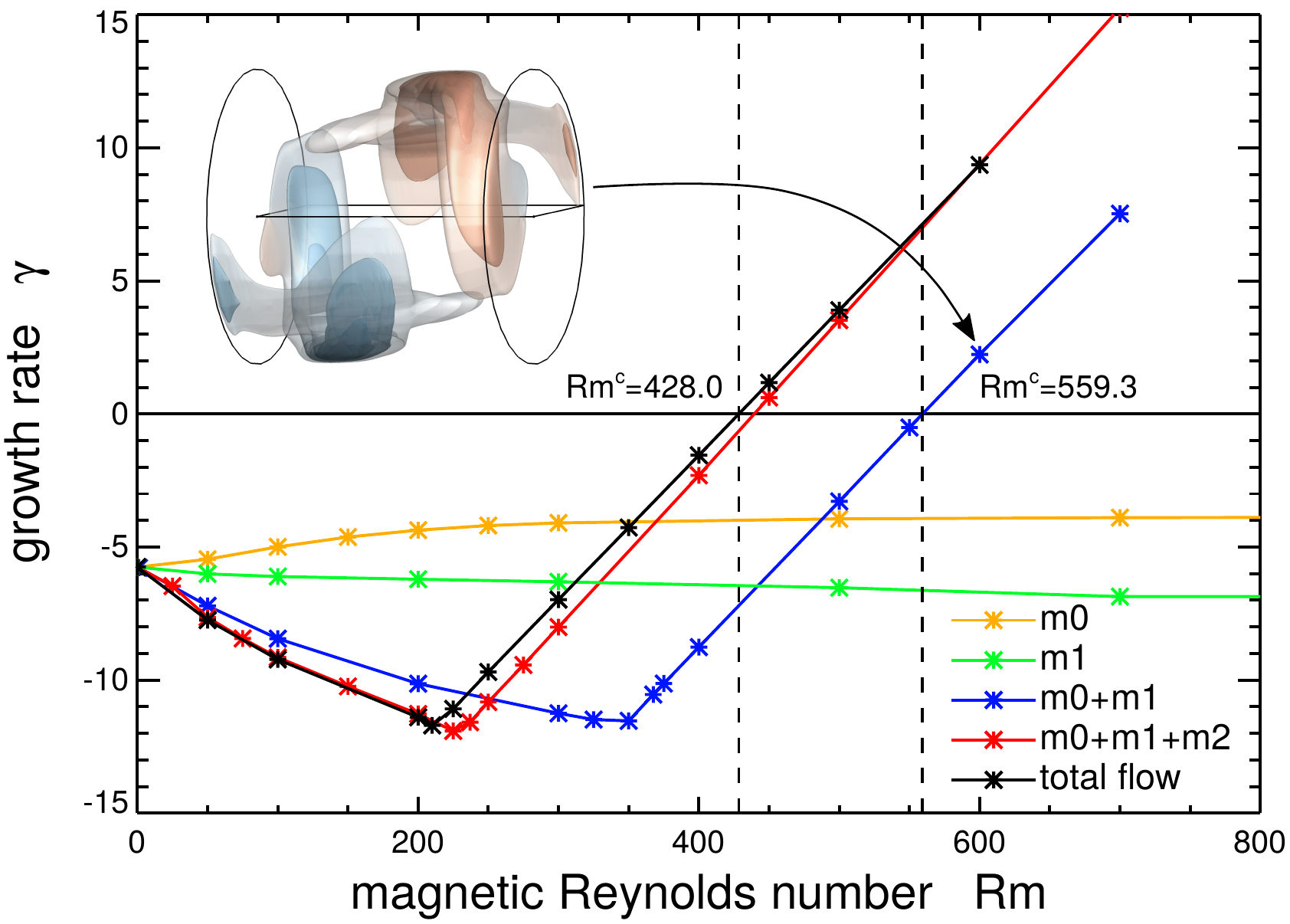}
\caption{ 
Growth rates for combinations of various azimuthal modes from the
velocity field obtained at simulations at ${\rm{Re}}=10^4$ and
${\rm{Po}}=0.1$.  The insert drawing depicts isosurfaces of the
magnetic energy density mapped with $B_{\varphi}$.  In the precession
frame the field structure propagates around the cylinder axis (see
movie at \cite{movie02}).
\label{fig::gr}} 
\end{figure}

In the following we use the velocity fields obtained from the
hydrodynamic simulations, validated by UDV flow measurements, as basis
for kinematic dynamo models. We concentrate on the strongly precessing
regime around ${\rm{Po}} \approx 0.1$ so that the flow is determined
by standing inertial waves which makes the time-averaged velocity
field appropriate for the application in kinematic simulations.  The
flow field is further decomposed into separate azimuthal modes
$m=0,1,2$ in order to carve out the impact of the individual
contributions on the dynamo.  The temporal evolution of the magnetic
flux density $\vec{B}$ induced by a given time-averaged flow
$\vec{\bar{u}}$ of a conducting liquid is determined by the induction
equation
\begin{equation}
\frac{\partial}{\partial t}\vec{B}
=\nabla\times\left(\vec{\bar{u}}\times\vec{B}
-\eta\nabla\times\vec{B}\right). 
\label{eq::magind}
\end{equation}
With the Ansatz $\vec{B}(\vec{r},t)=\vec{B}_0(\vec{r})e^{\sigma t}$
the solution of Eq.~(\ref{eq::magind}) is a linear problem with the
real part of the eigenvalue $\sigma$ representing the magnetic field
growth rate $\gamma$. We solve Eq.~(\ref{eq::magind}) numerically with
pseudo-vacuum boundary conditions for the magnetic field, and the
growth rates are computed from the time-evolution of the magnetic
field.  Except for the velocity field at ${\rm{Po}}=0.1$, the
kinematic models either show no dynamo or do so at best for
magnetic Reynolds numbers far above the values that will be attainable in the
planned dynamo experiment (e.g.  ${\rm{Rm}}^{\rm{c}}\approx 5000$ for
${\rm{Po}}=0.0875$).  Taking the time-averaged flow field from 
hydrodynamic simulations at ${\rm{Re}}=10^4$ and ${\rm{Po}}=0.1$, we
find dynamos at much reduced ${\rm{Rm}}$. The kinematic growth rates
for this particular case are shown in Fig.~\ref{fig::gr} where we
distinguish five different set-ups.  We find that neither the
axisymmetric flow ($m=0$, orange curve) nor the directly forced flow
($m=1$, green curve) alone are capable of driving a dynamo.
The latter was expected because the structure of the primary flow is
too simplistic for dynamo action \cite{giesecke2015a}.
The failure of the pure axisymmetric flow to drive a dynamo confirms
our previous assumption of the inappropriate relation of axisymmetric
poloidal and toroidal flow components.  However, when summing up both
contributions we obtain dynamo action at a critical magnetic Reynolds
number ${\rm{Rm}}^{\rm{c}}\approx 560$ (blue curve).  This value
decreases to ${\rm{Rm}}^{\rm{c}}\approx 430$ when further including
the $m=2$ modes (red curve), most probably because this contribution,
which is dominated by the $(m,k)=(2,2)$ mode, increases the breaking
of the equatorial symmetry, which is beneficial for
precession driven dynamos \cite{tilgner2005}.  Other
contributions with higher wave numbers are less important and no
significant further reduction of ${\rm{Rm}}^{\rm{c}}$ is obtained when
using the total time-averaged flow which yields
${\rm{Rm}}^{\rm{c}}\approx 428$ (black curve).

The water experiments indicate that the flow structure does not change
much when increasing ${\rm{Re}}$ \cite{supplement01}, albeit the
corresponding decrease of ${\rm{Po}}^{\rm{c}}$ does not follow a
simple scaling law (Fig.~\ref{fig::amp}c). However, it is known from
measurements of the internal pressure that the sudden drop of the
$m=1$ mode, which constitutes the second criteria for
${\rm{Po}}^{\rm{c}}$, only weakly depends on ${\rm{Re}}$ if ${\rm{Re}}
\gtrsim 5\times 10^5$ \cite{herault2015}.  This is already indicated
in our experiments when increasing ${\rm{Re}}$ from $4\times
10^4$ to $10^5$.  The width within which we observe the
axisymmetric mode ($\Delta {\rm{Po}}\approx 0.006$) corresponds nearly
exactly to the width of the hysteresis found in \cite{herault2015}
around ${\rm{Re}}\sim O(10^6)$ at a precession ratio comparable with
${\rm{Po}}^{\rm{c}}$ in our experiments at ${\rm{Re}}=10^5$. It seems
likely that both phenomena are closely connected, with the
$(m,k)=(0,2)$ mode being a precursor for the transition to the
turbulent state observed in \cite{herault2015}.  In the limit of large
${\rm{Re}}$ as they will occur in the liquid sodium experiment (up to
${\rm{Re}}\approx 10^8$), we thus expect dynamo action to arise in
connection with the non-geostrophic axisymmetric mode within a width
of $\Delta {\rm{Po}}\approx 0.006$ around ${\rm{Po}}$ not much smaller
than ${\rm{Po}}^{\rm{c}}$ in our experiments at ${\rm{Re}} = 10^5$.

Our results reveal a first promising -- though narrow -- regime,
defined by the presence of the axisymmetric mode, within which we
expect dynamo action in the planned dynamo experiment.  This is not a
turbulent dynamo since there is no significant amount of turbulence as
it would result, for example, from the resonant collapse reported in
experimental studies of precessing flows with small nutation angles
\cite{mcewan1970,*manasseh1994}.  Our model rather constitutes a
laminar dynamo driven by few large scale velocity modes, and our
simulations and measurements indicate that time-dependent
contributions remain weak even at the largest ${\rm{Re}}$ with the
spectra always being determined by standing inertial waves.  This is
in contrast to the flow in the VKS dynamo where the fluctuations are
of the same order as the mean flow. Instead, a comparison with the
Riga Dynamo is more appropriate, in which a fully developed turbulence
arises on top of a mean flow \cite{gailitis2008}, and calculations
based on the time-averaged flow field still provided good agreement
with the experiment \cite{gailitis2000,*gailitis2004} proving that the
turbulent $\beta$-effect remains negligible for such flows.

So far, we did not consider more realistic magnetic boundary
conditions, like an insulating outer domain or the finite conductivity
of the container made of stainless steel which will be focus of a
future study. Preliminary results 
from models including a thin outer layer with the electrical
conductivity reduced by a factor of 8 show an increment of
${\rm{Rm}}^{\rm{c}}$ by roughly $10\%$ which is still well within the
capabilities of the planned facility.

\acknowledgments{
This study has been conducted in the framework of the project DRESDYN
(DREsden Sodium facility for DYNamo and thermohydraulic studies) which
provides the platform for the precession dynamo experiment at
HZDR. The authors further acknowledge support by the Helmholtz Allianz
LIMTECH and Bernd Wustmann for the mechanical design of the experiment.}


\begin{thebibliography}{58}%
\makeatletter
\providecommand \@ifxundefined [1]{%
 \@ifx{#1\undefined}
}%
\providecommand \@ifnum [1]{%
 \ifnum #1\expandafter \@firstoftwo
 \else \expandafter \@secondoftwo
 \fi
}%
\providecommand \@ifx [1]{%
 \ifx #1\expandafter \@firstoftwo
 \else \expandafter \@secondoftwo
 \fi
}%
\providecommand \natexlab [1]{#1}%
\providecommand \enquote  [1]{``#1''}%
\providecommand \bibnamefont  [1]{#1}%
\providecommand \bibfnamefont [1]{#1}%
\providecommand \citenamefont [1]{#1}%
\providecommand \href@noop [0]{\@secondoftwo}%
\providecommand \href [0]{\begingroup \@sanitize@url \@href}%
\providecommand \@href[1]{\@@startlink{#1}\@@href}%
\providecommand \@@href[1]{\endgroup#1\@@endlink}%
\providecommand \@sanitize@url [0]{\catcode `\\12\catcode `\$12\catcode
  `\&12\catcode `\#12\catcode `\^12\catcode `\_12\catcode `\%12\relax}%
\providecommand \@@startlink[1]{}%
\providecommand \@@endlink[0]{}%
\providecommand \url  [0]{\begingroup\@sanitize@url \@url }%
\providecommand \@url [1]{\endgroup\@href {#1}{\urlprefix }}%
\providecommand \urlprefix  [0]{URL }%
\providecommand \Eprint [0]{\href }%
\providecommand \doibase [0]{http://dx.doi.org/}%
\providecommand \selectlanguage [0]{\@gobble}%
\providecommand \bibinfo  [0]{\@secondoftwo}%
\providecommand \bibfield  [0]{\@secondoftwo}%
\providecommand \translation [1]{[#1]}%
\providecommand \BibitemOpen [0]{}%
\providecommand \bibitemStop [0]{}%
\providecommand \bibitemNoStop [0]{.\EOS\space}%
\providecommand \EOS [0]{\spacefactor3000\relax}%
\providecommand \BibitemShut  [1]{\csname bibitem#1\endcsname}%
\let\auto@bib@innerbib\@empty
\bibitem [{\citenamefont {{Wicht}}\ and\ \citenamefont
  {{Tilgner}}(2010)}]{wicht2010}%
  \BibitemOpen
  \bibfield  {author} {\bibinfo {author} {\bibfnamefont {J.}~\bibnamefont
  {{Wicht}}}\ and\ \bibinfo {author} {\bibfnamefont {A.}~\bibnamefont
  {{Tilgner}}},\ }\href {\doibase 10.1007/s11214-010-9638-y} {\bibfield
  {journal} {\bibinfo  {journal} {\ssr}\ }\textbf {\bibinfo {volume} {152}},\
  \bibinfo {pages} {501} (\bibinfo {year} {2010})}\BibitemShut {NoStop}%
\bibitem [{\citenamefont {{Braginsky}}\ and\ \citenamefont
  {{Roberts}}(1995)}]{braginsky1995}%
  \BibitemOpen
  \bibfield  {author} {\bibinfo {author} {\bibfnamefont {S.~I.}\ \bibnamefont
  {{Braginsky}}}\ and\ \bibinfo {author} {\bibfnamefont {P.~H.}\ \bibnamefont
  {{Roberts}}},\ }\href {\doibase 10.1080/03091929508228992} {\bibfield
  {journal} {\bibinfo  {journal} {Geophys. Astrophys. Fluid Dyn.}\ }\textbf
  {\bibinfo {volume} {79}},\ \bibinfo {pages} {1} (\bibinfo {year}
  {1995})}\BibitemShut {NoStop}%
\bibitem [{\citenamefont {{Wu}}\ and\ \citenamefont
  {{Roberts}}(2013)}]{wu2013}%
  \BibitemOpen
  \bibfield  {author} {\bibinfo {author} {\bibfnamefont {C.~C.}\ \bibnamefont
  {{Wu}}}\ and\ \bibinfo {author} {\bibfnamefont {P.~H.}\ \bibnamefont
  {{Roberts}}},\ }\href {\doibase 10.1080/03091929.2012.682990} {\bibfield
  {journal} {\bibinfo  {journal} {Geophys. Astrophys. Fluid Dyn.}\ }\textbf
  {\bibinfo {volume} {107}},\ \bibinfo {pages} {20} (\bibinfo {year}
  {2013})}\BibitemShut {NoStop}%
\bibitem [{\citenamefont {{C{\'e}bron}}\ and\ \citenamefont
  {{Hollerbach}}(2014)}]{cebron2014}%
  \BibitemOpen
  \bibfield  {author} {\bibinfo {author} {\bibfnamefont {D.}~\bibnamefont
  {{C{\'e}bron}}}\ and\ \bibinfo {author} {\bibfnamefont {R.}~\bibnamefont
  {{Hollerbach}}},\ }\href {\doibase 10.1088/2041-8205/789/1/L25} {\bibfield
  {journal} {\bibinfo  {journal} {\apjl}\ }\textbf {\bibinfo {volume} {789}},\
  \bibinfo {eid} {L25} (\bibinfo {year} {2014})}\BibitemShut {NoStop}%
\bibitem [{\citenamefont {{Tilgner}}(2005)}]{tilgner2005}%
  \BibitemOpen
  \bibfield  {author} {\bibinfo {author} {\bibfnamefont {A.}~\bibnamefont
  {{Tilgner}}},\ }\href {\doibase 10.1063/1.1852576} {\bibfield  {journal}
  {\bibinfo  {journal} {\pof}\ }\textbf {\bibinfo {volume} {17}},\ \bibinfo
  {pages} {034104} (\bibinfo {year} {2005})}\BibitemShut {NoStop}%
\bibitem [{\citenamefont {{Dwyer}}\ \emph {et~al.}(2011)\citenamefont
  {{Dwyer}}, \citenamefont {{Stevenson}},\ and\ \citenamefont
  {{Nimmo}}}]{dwyer2011}%
  \BibitemOpen
  \bibfield  {author} {\bibinfo {author} {\bibfnamefont {C.~A.}\ \bibnamefont
  {{Dwyer}}}, \bibinfo {author} {\bibfnamefont {D.~J.}\ \bibnamefont
  {{Stevenson}}}, \ and\ \bibinfo {author} {\bibfnamefont {F.}~\bibnamefont
  {{Nimmo}}},\ }\href {\doibase 10.1038/nature10564} {\bibfield  {journal}
  {\bibinfo  {journal} {\nat}\ }\textbf {\bibinfo {volume} {479}},\ \bibinfo
  {pages} {212} (\bibinfo {year} {2011})}\BibitemShut {NoStop}%
\bibitem [{\citenamefont {{Noir}}\ and\ \citenamefont
  {{C{\'e}bron}}(2013)}]{noir2013}%
  \BibitemOpen
  \bibfield  {author} {\bibinfo {author} {\bibfnamefont {J.}~\bibnamefont
  {{Noir}}}\ and\ \bibinfo {author} {\bibfnamefont {D.}~\bibnamefont
  {{C{\'e}bron}}},\ }\href {\doibase 10.1017/jfm.2013.524} {\bibfield
  {journal} {\bibinfo  {journal} {\jfm}\ }\textbf {\bibinfo {volume} {737}},\
  \bibinfo {pages} {412} (\bibinfo {year} {2013})}\BibitemShut {NoStop}%
\bibitem [{\citenamefont {{Weiss}}\ and\ \citenamefont
  {{Tikoo}}(2014)}]{weiss2014}%
  \BibitemOpen
  \bibfield  {author} {\bibinfo {author} {\bibfnamefont {B.~P.}\ \bibnamefont
  {{Weiss}}}\ and\ \bibinfo {author} {\bibfnamefont {S.~M.}\ \bibnamefont
  {{Tikoo}}},\ }\href {\doibase 10.1126/science.1246753} {\bibfield  {journal}
  {\bibinfo  {journal} {Science}\ }\textbf {\bibinfo {volume} {346}},\ \bibinfo
  {pages} {1198} (\bibinfo {year} {2014})}\BibitemShut {NoStop}%
\bibitem [{\citenamefont {{Malkus}}(1968)}]{malkus1968}%
  \BibitemOpen
  \bibfield  {author} {\bibinfo {author} {\bibfnamefont {W.~V.~R.}\
  \bibnamefont {{Malkus}}},\ }\href {\doibase 10.1126/science.160.3825.259}
  {\bibfield  {journal} {\bibinfo  {journal} {Science}\ }\textbf {\bibinfo
  {volume} {160}},\ \bibinfo {pages} {259} (\bibinfo {year}
  {1968})}\BibitemShut {NoStop}%
\bibitem [{\citenamefont {{Vanyo}}(1991)}]{vanyo1991}%
  \BibitemOpen
  \bibfield  {author} {\bibinfo {author} {\bibfnamefont {J.~P.}\ \bibnamefont
  {{Vanyo}}},\ }\href {\doibase 10.1080/03091929108227780} {\bibfield
  {journal} {\bibinfo  {journal} {\gafd}\ }\textbf {\bibinfo {volume} {59}},\
  \bibinfo {pages} {209} (\bibinfo {year} {1991})}\BibitemShut {NoStop}%
\bibitem [{\citenamefont {{Gans}}(1970)}]{gans1970}%
  \BibitemOpen
  \bibfield  {author} {\bibinfo {author} {\bibfnamefont {R.~F.}\ \bibnamefont
  {{Gans}}},\ }\href {\doibase 10.1017/S0022112070000976} {\bibfield  {journal}
  {\bibinfo  {journal} {\jfm}\ }\textbf {\bibinfo {volume} {41}},\ \bibinfo
  {pages} {865} (\bibinfo {year} {1970})}\BibitemShut {NoStop}%
\bibitem [{\citenamefont {{Manasseh}}(1992)}]{manasseh1992}%
  \BibitemOpen
  \bibfield  {author} {\bibinfo {author} {\bibfnamefont {R.}~\bibnamefont
  {{Manasseh}}},\ }\href {\doibase 10.1017/S0022112092002726} {\bibfield
  {journal} {\bibinfo  {journal} {\jfm}\ }\textbf {\bibinfo {volume} {243}},\
  \bibinfo {pages} {261} (\bibinfo {year} {1992})}\BibitemShut {NoStop}%
\bibitem [{\citenamefont {{Manasseh}}(1996)}]{manasseh1996}%
  \BibitemOpen
  \bibfield  {author} {\bibinfo {author} {\bibfnamefont {R.}~\bibnamefont
  {{Manasseh}}},\ }\href {\doibase 10.1017/S0022112096002388} {\bibfield
  {journal} {\bibinfo  {journal} {\jfm}\ }\textbf {\bibinfo {volume} {315}},\
  \bibinfo {pages} {151} (\bibinfo {year} {1996})}\BibitemShut {NoStop}%
\bibitem [{\citenamefont {{Noir}}\ \emph {et~al.}(2003)\citenamefont {{Noir}},
  \citenamefont {{Cardin}}, \citenamefont {{Jault}},\ and\ \citenamefont
  {{Masson}}}]{noir2003}%
  \BibitemOpen
  \bibfield  {author} {\bibinfo {author} {\bibfnamefont {J.}~\bibnamefont
  {{Noir}}}, \bibinfo {author} {\bibfnamefont {P.}~\bibnamefont {{Cardin}}},
  \bibinfo {author} {\bibfnamefont {D.}~\bibnamefont {{Jault}}}, \ and\
  \bibinfo {author} {\bibfnamefont {J.-P.}\ \bibnamefont {{Masson}}},\ }\href
  {\doibase 10.1046/j.1365-246X.2003.01934.x} {\bibfield  {journal} {\bibinfo
  {journal} {\gji}\ }\textbf {\bibinfo {volume} {154}},\ \bibinfo {pages} {407}
  (\bibinfo {year} {2003})}\BibitemShut {NoStop}%
\bibitem [{\citenamefont {Lagrange}\ \emph {et~al.}(2011)\citenamefont
  {Lagrange}, \citenamefont {Meunier}, \citenamefont {Nadal},\ and\
  \citenamefont {Eloy}}]{lagrange2011}%
  \BibitemOpen
  \bibfield  {author} {\bibinfo {author} {\bibfnamefont {R.}~\bibnamefont
  {Lagrange}}, \bibinfo {author} {\bibfnamefont {P.}~\bibnamefont {Meunier}},
  \bibinfo {author} {\bibfnamefont {F.}~\bibnamefont {Nadal}}, \ and\ \bibinfo
  {author} {\bibfnamefont {C.}~\bibnamefont {Eloy}},\ }\href {\doibase
  10.1017/S0022112010004040} {\bibfield  {journal} {\bibinfo  {journal} {\jfm}\
  }\textbf {\bibinfo {volume} {666}},\ \bibinfo {pages} {104} (\bibinfo {year}
  {2011})}\BibitemShut {NoStop}%
\bibitem [{\citenamefont {{Goto}}\ \emph {et~al.}(2014)\citenamefont {{Goto}},
  \citenamefont {{Matsunaga}}, \citenamefont {{Fujiwara}}, \citenamefont
  {{Nishioka}}, \citenamefont {{Kida}}, \citenamefont {{Yamato}},\ and\
  \citenamefont {{Tsuda}}}]{goto2014}%
  \BibitemOpen
  \bibfield  {author} {\bibinfo {author} {\bibfnamefont {S.}~\bibnamefont
  {{Goto}}}, \bibinfo {author} {\bibfnamefont {A.}~\bibnamefont {{Matsunaga}}},
  \bibinfo {author} {\bibfnamefont {M.}~\bibnamefont {{Fujiwara}}}, \bibinfo
  {author} {\bibfnamefont {M.}~\bibnamefont {{Nishioka}}}, \bibinfo {author}
  {\bibfnamefont {S.}~\bibnamefont {{Kida}}}, \bibinfo {author} {\bibfnamefont
  {M.}~\bibnamefont {{Yamato}}}, \ and\ \bibinfo {author} {\bibfnamefont
  {S.}~\bibnamefont {{Tsuda}}},\ }\href {\doibase 10.1063/1.4874695} {\bibfield
   {journal} {\bibinfo  {journal} {\pof}\ }\textbf {\bibinfo {volume} {26}},\
  \bibinfo {eid} {055107} (\bibinfo {year} {2014})}\BibitemShut {NoStop}%
\bibitem [{\citenamefont {{Lin}}\ \emph {et~al.}(2014)\citenamefont {{Lin}},
  \citenamefont {{Noir}},\ and\ \citenamefont {{Jackson}}}]{lin2014}%
  \BibitemOpen
  \bibfield  {author} {\bibinfo {author} {\bibfnamefont {Y.}~\bibnamefont
  {{Lin}}}, \bibinfo {author} {\bibfnamefont {J.}~\bibnamefont {{Noir}}}, \
  and\ \bibinfo {author} {\bibfnamefont {A.}~\bibnamefont {{Jackson}}},\ }\href
  {\doibase 10.1063/1.4871026} {\bibfield  {journal} {\bibinfo  {journal}
  {\pof}\ }\textbf {\bibinfo {volume} {26}},\ \bibinfo {eid} {046604} (\bibinfo
  {year} {2014})}\BibitemShut {NoStop}%
\bibitem [{\citenamefont {{Lin}}\ \emph {et~al.}(2015)\citenamefont {{Lin}},
  \citenamefont {{Marti}},\ and\ \citenamefont {{Noir}}}]{lin2015}%
  \BibitemOpen
  \bibfield  {author} {\bibinfo {author} {\bibfnamefont {Y.}~\bibnamefont
  {{Lin}}}, \bibinfo {author} {\bibfnamefont {P.}~\bibnamefont {{Marti}}}, \
  and\ \bibinfo {author} {\bibfnamefont {J.}~\bibnamefont {{Noir}}},\ }\href
  {\doibase 10.1063/1.4916234} {\bibfield  {journal} {\bibinfo  {journal}
  {\pof}\ }\textbf {\bibinfo {volume} {27}},\ \bibinfo {eid} {046601} (\bibinfo
  {year} {2015})}\BibitemShut {NoStop}%
\bibitem [{\citenamefont {{Marques}}\ and\ \citenamefont
  {{Lopez}}(2015)}]{marques2015}%
  \BibitemOpen
  \bibfield  {author} {\bibinfo {author} {\bibfnamefont {F.}~\bibnamefont
  {{Marques}}}\ and\ \bibinfo {author} {\bibfnamefont {J.~M.}\ \bibnamefont
  {{Lopez}}},\ }\href {\doibase 10.1017/jfm.2015.524} {\bibfield  {journal}
  {\bibinfo  {journal} {\jfm}\ }\textbf {\bibinfo {volume} {782}},\ \bibinfo
  {pages} {63} (\bibinfo {year} {2015})}\BibitemShut {NoStop}%
\bibitem [{\citenamefont {{Giesecke}}\ \emph
  {et~al.}(2015{\natexlab{a}})\citenamefont {{Giesecke}}, \citenamefont
  {{Albrecht}}, \citenamefont {{Gundrum}}, \citenamefont {{Herault}},\ and\
  \citenamefont {{Stefani}}}]{giesecke2015b}%
  \BibitemOpen
  \bibfield  {author} {\bibinfo {author} {\bibfnamefont {A.}~\bibnamefont
  {{Giesecke}}}, \bibinfo {author} {\bibfnamefont {T.}~\bibnamefont
  {{Albrecht}}}, \bibinfo {author} {\bibfnamefont {T.}~\bibnamefont
  {{Gundrum}}}, \bibinfo {author} {\bibfnamefont {J.}~\bibnamefont
  {{Herault}}}, \ and\ \bibinfo {author} {\bibfnamefont {F.}~\bibnamefont
  {{Stefani}}},\ }\href {\doibase 10.1088/1367-2630/17/11/113044} {\bibfield
  {journal} {\bibinfo  {journal} {\njp}\ }\textbf {\bibinfo {volume} {17}},\
  \bibinfo {eid} {113044} (\bibinfo {year} {2015}{\natexlab{a}})}\BibitemShut
  {NoStop}%
\bibitem [{\citenamefont {{Lopez}}\ and\ \citenamefont
  {{Marques}}(2016)}]{lopez2016}%
  \BibitemOpen
  \bibfield  {author} {\bibinfo {author} {\bibfnamefont {J.~M.}\ \bibnamefont
  {{Lopez}}}\ and\ \bibinfo {author} {\bibfnamefont {F.}~\bibnamefont
  {{Marques}}},\ }\href {\doibase 10.1103/PhysRevFluids.1.023602} {\bibfield
  {journal} {\bibinfo  {journal} {\prf}\ }\textbf {\bibinfo {volume} {1}},\
  \bibinfo {eid} {023602} (\bibinfo {year} {2016})}\BibitemShut {NoStop}%
\bibitem [{\citenamefont {{Mouhali}}\ \emph {et~al.}(2012)\citenamefont
  {{Mouhali}}, \citenamefont {{Lehner}}, \citenamefont {{L{\'e}orat}},\ and\
  \citenamefont {{Vitry}}}]{mouhali2012}%
  \BibitemOpen
  \bibfield  {author} {\bibinfo {author} {\bibfnamefont {W.}~\bibnamefont
  {{Mouhali}}}, \bibinfo {author} {\bibfnamefont {T.}~\bibnamefont {{Lehner}}},
  \bibinfo {author} {\bibfnamefont {J.}~\bibnamefont {{L{\'e}orat}}}, \ and\
  \bibinfo {author} {\bibfnamefont {R.}~\bibnamefont {{Vitry}}},\ }\href
  {\doibase 10.1007/s00348-012-1385-2} {\bibfield  {journal} {\bibinfo
  {journal} {Exp. Fluids}\ }\textbf {\bibinfo {volume} {53}},\ \bibinfo {pages}
  {1693} (\bibinfo {year} {2012})}\BibitemShut {NoStop}%
\bibitem [{\citenamefont {{Le Bars}}\ \emph {et~al.}(2015)\citenamefont {{Le
  Bars}}, \citenamefont {{C{\'e}bron}},\ and\ \citenamefont {{Le
  Gal}}}]{lebars2015}%
  \BibitemOpen
  \bibfield  {author} {\bibinfo {author} {\bibfnamefont {M.}~\bibnamefont {{Le
  Bars}}}, \bibinfo {author} {\bibfnamefont {D.}~\bibnamefont {{C{\'e}bron}}},
  \ and\ \bibinfo {author} {\bibfnamefont {P.}~\bibnamefont {{Le Gal}}},\
  }\href {\doibase 10.1146/annurev-fluid-010814-014556} {\bibfield  {journal}
  {\bibinfo  {journal} {Annu. Rev. Fluid Mech.}\ }\textbf {\bibinfo {volume}
  {47}},\ \bibinfo {pages} {163} (\bibinfo {year} {2015})}\BibitemShut
  {NoStop}%
\bibitem [{\citenamefont {L{\'e}orat}\ \emph {et~al.}(2003)\citenamefont
  {L{\'e}orat}, \citenamefont {Rigaud}, \citenamefont {Vitry},\ and\
  \citenamefont {Herpe}}]{leorat2003}%
  \BibitemOpen
  \bibfield  {author} {\bibinfo {author} {\bibfnamefont {J.}~\bibnamefont
  {L{\'e}orat}}, \bibinfo {author} {\bibfnamefont {F.}~\bibnamefont {Rigaud}},
  \bibinfo {author} {\bibfnamefont {R.}~\bibnamefont {Vitry}}, \ and\ \bibinfo
  {author} {\bibfnamefont {G.}~\bibnamefont {Herpe}},\ }\href@noop {}
  {\bibfield  {journal} {\bibinfo  {journal} {Magnetohydrodynamics}\ }\textbf
  {\bibinfo {volume} {39}},\ \bibinfo {pages} {321} (\bibinfo {year}
  {2003})}\BibitemShut {NoStop}%
\bibitem [{\citenamefont {{L{\'e}orat}}(2006)}]{leorat2006}%
  \BibitemOpen
  \bibfield  {author} {\bibinfo {author} {\bibfnamefont {J.}~\bibnamefont
  {{L{\'e}orat}}},\ }\href@noop {} {\bibfield  {journal} {\bibinfo  {journal}
  {Magnetohydrodynamics}\ }\textbf {\bibinfo {volume} {42}},\ \bibinfo {pages}
  {143} (\bibinfo {year} {2006})}\BibitemShut {NoStop}%
\bibitem [{\citenamefont {{Stefani}}\ \emph {et~al.}(2008)\citenamefont
  {{Stefani}}, \citenamefont {{Gailitis}},\ and\ \citenamefont
  {{Gerbeth}}}]{stefani2008}%
  \BibitemOpen
  \bibfield  {author} {\bibinfo {author} {\bibfnamefont {F.}~\bibnamefont
  {{Stefani}}}, \bibinfo {author} {\bibfnamefont {A.}~\bibnamefont
  {{Gailitis}}}, \ and\ \bibinfo {author} {\bibfnamefont {G.}~\bibnamefont
  {{Gerbeth}}},\ }\href {\doibase 10.1002/zamm.200800102} {\bibfield  {journal}
  {\bibinfo  {journal} {Zeitschrift f{\"u}r Angewandte Mathematik und
  Mechanik}\ }\textbf {\bibinfo {volume} {88}},\ \bibinfo {pages} {930}
  (\bibinfo {year} {2008})}\BibitemShut {NoStop}%
\bibitem [{\citenamefont {{Stefani}}\ \emph {et~al.}(2012)\citenamefont
  {{Stefani}}, \citenamefont {{Eckert}}, \citenamefont {{Gerbeth}},
  \citenamefont {{Giesecke}}, \citenamefont {{Gundrum}}, \citenamefont
  {{Steglich}}, \citenamefont {{Weier}},\ and\ \citenamefont
  {{Wustmann}}}]{stefani2012}%
  \BibitemOpen
  \bibfield  {author} {\bibinfo {author} {\bibfnamefont {F.}~\bibnamefont
  {{Stefani}}}, \bibinfo {author} {\bibfnamefont {S.}~\bibnamefont {{Eckert}}},
  \bibinfo {author} {\bibfnamefont {G.}~\bibnamefont {{Gerbeth}}}, \bibinfo
  {author} {\bibfnamefont {A.}~\bibnamefont {{Giesecke}}}, \bibinfo {author}
  {\bibfnamefont {T.}~\bibnamefont {{Gundrum}}}, \bibinfo {author}
  {\bibfnamefont {C.}~\bibnamefont {{Steglich}}}, \bibinfo {author}
  {\bibfnamefont {T.}~\bibnamefont {{Weier}}}, \ and\ \bibinfo {author}
  {\bibfnamefont {B.}~\bibnamefont {{Wustmann}}},\ }\href@noop {} {\bibfield
  {journal} {\bibinfo  {journal} {Magnetohydrodynamics}\ }\textbf {\bibinfo
  {volume} {48}},\ \bibinfo {pages} {103} (\bibinfo {year} {2012})}\BibitemShut
  {NoStop}%
\bibitem [{\citenamefont {{Gans}}(1971)}]{gans1971}%
  \BibitemOpen
  \bibfield  {author} {\bibinfo {author} {\bibfnamefont {R.~F.}\ \bibnamefont
  {{Gans}}},\ }\href {\doibase 10.1017/S0022112071003021} {\bibfield  {journal}
  {\bibinfo  {journal} {\jfm}\ }\textbf {\bibinfo {volume} {45}},\ \bibinfo
  {pages} {111} (\bibinfo {year} {1971})}\BibitemShut {NoStop}%
\bibitem [{\citenamefont {{Wu}}\ and\ \citenamefont
  {{Roberts}}(2009)}]{wu2009}%
  \BibitemOpen
  \bibfield  {author} {\bibinfo {author} {\bibfnamefont {C.-C.}\ \bibnamefont
  {{Wu}}}\ and\ \bibinfo {author} {\bibfnamefont {P.}~\bibnamefont
  {{Roberts}}},\ }\href {\doibase 10.1080/03091920903311788} {\bibfield
  {journal} {\bibinfo  {journal} {\gafd}\ }\textbf {\bibinfo {volume} {103}},\
  \bibinfo {pages} {467} (\bibinfo {year} {2009})}\BibitemShut {NoStop}%
\bibitem [{\citenamefont {Ernst-Hullermann}\ \emph {et~al.}(2011)\citenamefont
  {Ernst-Hullermann}, \citenamefont {Harder},\ and\ \citenamefont
  {Hansen}}]{hullermann2011}%
  \BibitemOpen
  \bibfield  {author} {\bibinfo {author} {\bibfnamefont {J.}~\bibnamefont
  {Ernst-Hullermann}}, \bibinfo {author} {\bibfnamefont {H.}~\bibnamefont
  {Harder}}, \ and\ \bibinfo {author} {\bibfnamefont {U.}~\bibnamefont
  {Hansen}},\ }\href@noop {} {\bibfield  {journal} {\bibinfo  {journal}
  {Geophys. J. Int.}\ }\textbf {\bibinfo {volume} {195}},\ \bibinfo {pages}
  {1395} (\bibinfo {year} {2011})}\BibitemShut {NoStop}%
\bibitem [{\citenamefont {{Lin}}\ \emph {et~al.}(2016)\citenamefont {{Lin}},
  \citenamefont {{Marti}}, \citenamefont {{Noir}},\ and\ \citenamefont
  {{Jackson}}}]{lin2016}%
  \BibitemOpen
  \bibfield  {author} {\bibinfo {author} {\bibfnamefont {Y.}~\bibnamefont
  {{Lin}}}, \bibinfo {author} {\bibfnamefont {P.}~\bibnamefont {{Marti}}},
  \bibinfo {author} {\bibfnamefont {J.}~\bibnamefont {{Noir}}}, \ and\ \bibinfo
  {author} {\bibfnamefont {A.}~\bibnamefont {{Jackson}}},\ }\href {\doibase
  10.1063/1.4954295} {\bibfield  {journal} {\bibinfo  {journal} {\pof}\
  }\textbf {\bibinfo {volume} {28}},\ \bibinfo {eid} {066601} (\bibinfo {year}
  {2016})}\BibitemShut {NoStop}%
\bibitem [{\citenamefont {{Giesecke}}\ \emph
  {et~al.}(2015{\natexlab{b}})\citenamefont {{Giesecke}}, \citenamefont
  {{Albrecht}}, \citenamefont {{Gerbeth}}, \citenamefont {{Gundrum}},\ and\
  \citenamefont {{Stefani}}}]{giesecke2015a}%
  \BibitemOpen
  \bibfield  {author} {\bibinfo {author} {\bibfnamefont {A.}~\bibnamefont
  {{Giesecke}}}, \bibinfo {author} {\bibfnamefont {T.}~\bibnamefont
  {{Albrecht}}}, \bibinfo {author} {\bibfnamefont {G.}~\bibnamefont
  {{Gerbeth}}}, \bibinfo {author} {\bibfnamefont {T.}~\bibnamefont
  {{Gundrum}}}, \ and\ \bibinfo {author} {\bibfnamefont {F.}~\bibnamefont
  {{Stefani}}},\ }\href@noop {} {\bibfield  {journal} {\bibinfo  {journal}
  {Magnetohydrodynamics}\ }\textbf {\bibinfo {volume} {51}},\ \bibinfo {pages}
  {293} (\bibinfo {year} {2015}{\natexlab{b}})}\BibitemShut {NoStop}%
\bibitem [{\citenamefont {{Nore}}\ \emph {et~al.}(2011)\citenamefont {{Nore}},
  \citenamefont {{L{\'e}orat}}, \citenamefont {{Guermond}},\ and\ \citenamefont
  {{Luddens}}}]{nore2011}%
  \BibitemOpen
  \bibfield  {author} {\bibinfo {author} {\bibfnamefont {C.}~\bibnamefont
  {{Nore}}}, \bibinfo {author} {\bibfnamefont {J.}~\bibnamefont
  {{L{\'e}orat}}}, \bibinfo {author} {\bibfnamefont {J.-L.}\ \bibnamefont
  {{Guermond}}}, \ and\ \bibinfo {author} {\bibfnamefont {F.}~\bibnamefont
  {{Luddens}}},\ }\href {\doibase 10.1103/PhysRevE.84.016317} {\bibfield
  {journal} {\bibinfo  {journal} {\pre}\ }\textbf {\bibinfo {volume} {84}},\
  \bibinfo {eid} {016317} (\bibinfo {year} {2011})}\BibitemShut {NoStop}%
\bibitem [{\citenamefont {{Goepfert}}\ and\ \citenamefont
  {{Tilgner}}(2016)}]{goepfert2016}%
  \BibitemOpen
  \bibfield  {author} {\bibinfo {author} {\bibfnamefont {O.}~\bibnamefont
  {{Goepfert}}}\ and\ \bibinfo {author} {\bibfnamefont {A.}~\bibnamefont
  {{Tilgner}}},\ }\href {\doibase 10.1088/1367-2630/18/10/103019} {\bibfield
  {journal} {\bibinfo  {journal} {\njp}\ }\textbf {\bibinfo {volume} {18}},\
  \bibinfo {eid} {103019} (\bibinfo {year} {2016})}\BibitemShut {NoStop}%
\bibitem [{mov({\natexlab{a}})}]{movie01}%
  \BibitemOpen
  \href@noop {} {} ({\natexlab{a}}),\ \bibinfo {note} {see Supplemental
  Material at [URL will be inserted by publisher] for animation of the
  three-dimensional structure of the axial flow at ${\rm{Re}}=10^4$ and
  ${\rm{Po}}=0.1$.}\BibitemShut {Stop}%
\bibitem [{\citenamefont {{Blackburn}}\ and\ \citenamefont
  {{Sherwin}}(2004)}]{blackburn2004}%
  \BibitemOpen
  \bibfield  {author} {\bibinfo {author} {\bibfnamefont {H.~M.}\ \bibnamefont
  {{Blackburn}}}\ and\ \bibinfo {author} {\bibfnamefont {S.~J.}\ \bibnamefont
  {{Sherwin}}},\ }\href {\doibase 10.1016/j.jcp.2004.02.013} {\bibfield
  {journal} {\bibinfo  {journal} {J. Comp. Phys.}\ }\textbf {\bibinfo {volume}
  {197}},\ \bibinfo {pages} {759} (\bibinfo {year} {2004})}\BibitemShut
  {NoStop}%
\bibitem [{\citenamefont {{Tilgner}}(1998)}]{tilgner1998}%
  \BibitemOpen
  \bibfield  {author} {\bibinfo {author} {\bibfnamefont {A.}~\bibnamefont
  {{Tilgner}}},\ }\href@noop {} {\bibfield  {journal} {\bibinfo  {journal}
  {Studia geoph. et geod.}\ }\textbf {\bibinfo {volume} {42}},\ \bibinfo
  {pages} {232} (\bibinfo {year} {1998})}\BibitemShut {NoStop}%
\bibitem [{\citenamefont {{Takeda}}(1986)}]{takeda1986}%
  \BibitemOpen
  \bibfield  {author} {\bibinfo {author} {\bibfnamefont {Y.}~\bibnamefont
  {{Takeda}}},\ }\href {\doibase 10.1016/0142-727X(86)90011-1} {\bibfield
  {journal} {\bibinfo  {journal} {Int. J. Heat Fluid Flow}\ }\textbf {\bibinfo
  {volume} {7}},\ \bibinfo {pages} {313 } (\bibinfo {year} {1986})}\BibitemShut
  {NoStop}%
\bibitem [{\citenamefont {{Vogt}}\ \emph {et~al.}(2014)\citenamefont {{Vogt}},
  \citenamefont {{R{\"a}biger}},\ and\ \citenamefont {{Eckert}}}]{vogt2014}%
  \BibitemOpen
  \bibfield  {author} {\bibinfo {author} {\bibfnamefont {T.}~\bibnamefont
  {{Vogt}}}, \bibinfo {author} {\bibfnamefont {D.}~\bibnamefont
  {{R{\"a}biger}}}, \ and\ \bibinfo {author} {\bibfnamefont {S.}~\bibnamefont
  {{Eckert}}},\ }\href {\doibase 10.1017/jfm.2014.371} {\bibfield  {journal}
  {\bibinfo  {journal} {\jfm}\ }\textbf {\bibinfo {volume} {753}},\ \bibinfo
  {pages} {472} (\bibinfo {year} {2014})}\BibitemShut {NoStop}%
\bibitem [{\citenamefont {{Thomson}}(1880)}]{kelvin1880}%
  \BibitemOpen
  \bibfield  {author} {\bibinfo {author} {\bibfnamefont {W.}~\bibnamefont
  {{Thomson}}},\ }\href@noop {} {\bibfield  {journal} {\bibinfo  {journal}
  {Philos. Mag.}\ }\textbf {\bibinfo {volume} {10}},\ \bibinfo {pages} {152}
  (\bibinfo {year} {1880})}\BibitemShut {NoStop}%
\bibitem [{\citenamefont {Greenspan}(1968)}]{greenspan}%
  \BibitemOpen
  \bibfield  {author} {\bibinfo {author} {\bibfnamefont {H.~P.}\ \bibnamefont
  {Greenspan}},\ }\href@noop {} {\emph {\bibinfo {title} {The theory of
  rotating fluids}}}\ (\bibinfo  {publisher} {Cambridge University Press},\
  \bibinfo {year} {1968})\BibitemShut {NoStop}%
\bibitem [{\citenamefont {{Liao}}\ and\ \citenamefont
  {{Zhang}}(2012)}]{liao2012}%
  \BibitemOpen
  \bibfield  {author} {\bibinfo {author} {\bibfnamefont {X.}~\bibnamefont
  {{Liao}}}\ and\ \bibinfo {author} {\bibfnamefont {K.}~\bibnamefont
  {{Zhang}}},\ }\href {\doibase 10.1017/jfm.2012.355} {\bibfield  {journal}
  {\bibinfo  {journal} {\jfm}\ }\textbf {\bibinfo {volume} {709}},\ \bibinfo
  {pages} {610} (\bibinfo {year} {2012})}\BibitemShut {NoStop}%
\bibitem [{\citenamefont {{Meunier}}\ \emph {et~al.}(2008)\citenamefont
  {{Meunier}}, \citenamefont {{Eloy}}, \citenamefont {{Lagrange}},\ and\
  \citenamefont {{Nadal}}}]{meunier2008}%
  \BibitemOpen
  \bibfield  {author} {\bibinfo {author} {\bibfnamefont {P.}~\bibnamefont
  {{Meunier}}}, \bibinfo {author} {\bibfnamefont {C.}~\bibnamefont {{Eloy}}},
  \bibinfo {author} {\bibfnamefont {R.}~\bibnamefont {{Lagrange}}}, \ and\
  \bibinfo {author} {\bibfnamefont {F.}~\bibnamefont {{Nadal}}},\ }\href
  {\doibase 10.1017/S0022112008000335} {\bibfield  {journal} {\bibinfo
  {journal} {\jfm}\ }\textbf {\bibinfo {volume} {599}},\ \bibinfo {pages} {405}
  (\bibinfo {year} {2008})}\BibitemShut {NoStop}%
\bibitem [{\citenamefont {{Monchaux}}\ \emph {et~al.}(2007)\citenamefont
  {{Monchaux}}, \citenamefont {{Berhanu}}, \citenamefont {{Bourgoin}},
  \citenamefont {{Moulin}}, \citenamefont {{Odier}}, \citenamefont {{Pinton}},
  \citenamefont {{Volk}}, \citenamefont {{Fauve}}, \citenamefont {{Mordant}},
  \citenamefont {{P{\'e}tr{\'e}lis}}, \citenamefont {{Chiffaudel}},
  \citenamefont {{Daviaud}}, \citenamefont {{Dubrulle}}, \citenamefont
  {{Gasquet}}, \citenamefont {{Mari{\'e}}},\ and\ \citenamefont
  {{Ravelet}}}]{monchaux2007}%
  \BibitemOpen
  \bibfield  {author} {\bibinfo {author} {\bibfnamefont {R.}~\bibnamefont
  {{Monchaux}}}, \bibinfo {author} {\bibfnamefont {M.}~\bibnamefont
  {{Berhanu}}}, \bibinfo {author} {\bibfnamefont {M.}~\bibnamefont
  {{Bourgoin}}}, \bibinfo {author} {\bibfnamefont {M.}~\bibnamefont
  {{Moulin}}}, \bibinfo {author} {\bibfnamefont {P.}~\bibnamefont {{Odier}}},
  \bibinfo {author} {\bibfnamefont {J.-F.}\ \bibnamefont {{Pinton}}}, \bibinfo
  {author} {\bibfnamefont {R.}~\bibnamefont {{Volk}}}, \bibinfo {author}
  {\bibfnamefont {S.}~\bibnamefont {{Fauve}}}, \bibinfo {author} {\bibfnamefont
  {N.}~\bibnamefont {{Mordant}}}, \bibinfo {author} {\bibfnamefont
  {F.}~\bibnamefont {{P{\'e}tr{\'e}lis}}}, \bibinfo {author} {\bibfnamefont
  {A.}~\bibnamefont {{Chiffaudel}}}, \bibinfo {author} {\bibfnamefont
  {F.}~\bibnamefont {{Daviaud}}}, \bibinfo {author} {\bibfnamefont
  {B.}~\bibnamefont {{Dubrulle}}}, \bibinfo {author} {\bibfnamefont
  {C.}~\bibnamefont {{Gasquet}}}, \bibinfo {author} {\bibfnamefont
  {L.}~\bibnamefont {{Mari{\'e}}}}, \ and\ \bibinfo {author} {\bibfnamefont
  {F.}~\bibnamefont {{Ravelet}}},\ }\href {\doibase
  10.1103/PhysRevLett.98.044502} {\bibfield  {journal} {\bibinfo  {journal}
  {\prl}\ }\textbf {\bibinfo {volume} {98}},\ \bibinfo {eid} {044502} (\bibinfo
  {year} {2007})}\BibitemShut {NoStop}%
\bibitem [{\citenamefont {{Dudley}}\ and\ \citenamefont
  {{James}}(1989)}]{dudley1989}%
  \BibitemOpen
  \bibfield  {author} {\bibinfo {author} {\bibfnamefont {M.~L.}\ \bibnamefont
  {{Dudley}}}\ and\ \bibinfo {author} {\bibfnamefont {R.~W.}\ \bibnamefont
  {{James}}},\ }\href {\doibase 10.1098/rspa.1989.0112} {\bibfield  {journal}
  {\bibinfo  {journal} {Proc. R. Soc. Lond. A}\ }\textbf {\bibinfo {volume}
  {425}},\ \bibinfo {pages} {407} (\bibinfo {year} {1989})}\BibitemShut
  {NoStop}%
\bibitem [{\citenamefont {{Ravelet}}\ \emph {et~al.}(2005)\citenamefont
  {{Ravelet}}, \citenamefont {{Chiffaudel}}, \citenamefont {{Daviaud}},\ and\
  \citenamefont {{L{\'e}orat}}}]{ravelet2005}%
  \BibitemOpen
  \bibfield  {author} {\bibinfo {author} {\bibfnamefont {F.}~\bibnamefont
  {{Ravelet}}}, \bibinfo {author} {\bibfnamefont {A.}~\bibnamefont
  {{Chiffaudel}}}, \bibinfo {author} {\bibfnamefont {F.}~\bibnamefont
  {{Daviaud}}}, \ and\ \bibinfo {author} {\bibfnamefont {J.}~\bibnamefont
  {{L{\'e}orat}}},\ }\href {\doibase 10.1063/1.2130745} {\bibfield  {journal}
  {\bibinfo  {journal} {\pof}\ }\textbf {\bibinfo {volume} {17}},\ \bibinfo
  {pages} {117104} (\bibinfo {year} {2005})}\BibitemShut {NoStop}%
\bibitem [{\citenamefont {{Greenspan}}(1969)}]{greenspan1969}%
  \BibitemOpen
  \bibfield  {author} {\bibinfo {author} {\bibfnamefont {H.~P.}\ \bibnamefont
  {{Greenspan}}},\ }\href {\doibase 10.1017/S0022112069001649} {\bibfield
  {journal} {\bibinfo  {journal} {\jfm}\ }\textbf {\bibinfo {volume} {36}},\
  \bibinfo {pages} {257} (\bibinfo {year} {1969})}\BibitemShut {NoStop}%
\bibitem [{\citenamefont {{Busse}}(1968)}]{busse1968}%
  \BibitemOpen
  \bibfield  {author} {\bibinfo {author} {\bibfnamefont {F.~H.}\ \bibnamefont
  {{Busse}}},\ }\href {\doibase 10.1017/S0022112068001655} {\bibfield
  {journal} {\bibinfo  {journal} {\jfm}\ }\textbf {\bibinfo {volume} {33}},\
  \bibinfo {pages} {739} (\bibinfo {year} {1968})}\BibitemShut {NoStop}%
\bibitem [{\citenamefont {{Tilgner}}(2007)}]{tilgner2007b}%
  \BibitemOpen
  \bibfield  {author} {\bibinfo {author} {\bibfnamefont {A.}~\bibnamefont
  {{Tilgner}}},\ }\href {\doibase 10.1103/PhysRevLett.99.194501} {\bibfield
  {journal} {\bibinfo  {journal} {Phys. Rev. Lett.}\ }\textbf {\bibinfo
  {volume} {99}},\ \bibinfo {pages} {194501} (\bibinfo {year}
  {2007})}\BibitemShut {NoStop}%
\bibitem [{mov({\natexlab{b}})}]{movie02}%
  \BibitemOpen
  \href@noop {} {} ({\natexlab{b}}),\ \bibinfo {note} {see Supplemental
  Material at [URL will be inserted by publisher] for animation of the
  three-dimensional structure of the magnetic energy density mapped with the
  azimuthal field.}\BibitemShut {Stop}%
\bibitem [{sup()}]{supplement01}%
  \BibitemOpen
  \href@noop {} {}\bibinfo {note} {See Supplemental Material at [URL will be
  inserted by publisher] for flow structure from UDV measurements up to
  ${\rm{Re}}=10^5$.}\BibitemShut {Stop}%
\bibitem [{\citenamefont {{Herault}}\ \emph {et~al.}(2015)\citenamefont
  {{Herault}}, \citenamefont {{Gundrum}}, \citenamefont {{Giesecke}},\ and\
  \citenamefont {{Stefani}}}]{herault2015}%
  \BibitemOpen
  \bibfield  {author} {\bibinfo {author} {\bibfnamefont {J.}~\bibnamefont
  {{Herault}}}, \bibinfo {author} {\bibfnamefont {T.}~\bibnamefont
  {{Gundrum}}}, \bibinfo {author} {\bibfnamefont {A.}~\bibnamefont
  {{Giesecke}}}, \ and\ \bibinfo {author} {\bibfnamefont {F.}~\bibnamefont
  {{Stefani}}},\ }\href {\doibase 10.1063/1.4936653} {\bibfield  {journal}
  {\bibinfo  {journal} {\pof}\ }\textbf {\bibinfo {volume} {27}},\ \bibinfo
  {eid} {124102} (\bibinfo {year} {2015})}\BibitemShut {NoStop}%
\bibitem [{\citenamefont {{McEwan}}(1970)}]{mcewan1970}%
  \BibitemOpen
  \bibfield  {author} {\bibinfo {author} {\bibfnamefont {A.~D.}\ \bibnamefont
  {{McEwan}}},\ }\href {\doibase 10.1017/S0022112070000344} {\bibfield
  {journal} {\bibinfo  {journal} {\jfm}\ }\textbf {\bibinfo {volume} {40}},\
  \bibinfo {pages} {603} (\bibinfo {year} {1970})}\BibitemShut {NoStop}%
\bibitem [{\citenamefont {{Manasseh}}(1994)}]{manasseh1994}%
  \BibitemOpen
  \bibfield  {author} {\bibinfo {author} {\bibfnamefont {R.}~\bibnamefont
  {{Manasseh}}},\ }\href {\doibase 10.1017/S0022112094000868} {\bibfield
  {journal} {\bibinfo  {journal} {\jfm}\ }\textbf {\bibinfo {volume} {265}},\
  \bibinfo {pages} {345} (\bibinfo {year} {1994})}\BibitemShut {NoStop}%
\bibitem [{\citenamefont {{Gailitis}}\ \emph {et~al.}(2008)\citenamefont
  {{Gailitis}}, \citenamefont {{Gerbeth}}, \citenamefont {{Gundrum}},
  \citenamefont {{Lielausis}}, \citenamefont {{Platacis}},\ and\ \citenamefont
  {{Stefani}}}]{gailitis2008}%
  \BibitemOpen
  \bibfield  {author} {\bibinfo {author} {\bibfnamefont {A.}~\bibnamefont
  {{Gailitis}}}, \bibinfo {author} {\bibfnamefont {G.}~\bibnamefont
  {{Gerbeth}}}, \bibinfo {author} {\bibfnamefont {T.}~\bibnamefont
  {{Gundrum}}}, \bibinfo {author} {\bibfnamefont {O.}~\bibnamefont
  {{Lielausis}}}, \bibinfo {author} {\bibfnamefont {E.}~\bibnamefont
  {{Platacis}}}, \ and\ \bibinfo {author} {\bibfnamefont {F.}~\bibnamefont
  {{Stefani}}},\ }\href {\doibase 10.1016/j.crhy.2008.07.004} {\bibfield
  {journal} {\bibinfo  {journal} {C. R. Phys.}\ }\textbf {\bibinfo {volume}
  {9}},\ \bibinfo {pages} {721} (\bibinfo {year} {2008})},\ \Eprint
  {http://arxiv.org/abs/0807.0305} {0807.0305} \BibitemShut {NoStop}%
\bibitem [{\citenamefont {{Gailitis}}\ \emph {et~al.}(2000)\citenamefont
  {{Gailitis}}, \citenamefont {{Lielausis}}, \citenamefont {{Dement'ev}},
  \citenamefont {{Platacis}}, \citenamefont {{Cifersons}}, \citenamefont
  {{Gerbeth}}, \citenamefont {{Gundrum}}, \citenamefont {{Stefani}},
  \citenamefont {{Christen}}, \citenamefont {{H{\"a}nel}},\ and\ \citenamefont
  {{Will}}}]{gailitis2000}%
  \BibitemOpen
  \bibfield  {author} {\bibinfo {author} {\bibfnamefont {A.}~\bibnamefont
  {{Gailitis}}}, \bibinfo {author} {\bibfnamefont {O.}~\bibnamefont
  {{Lielausis}}}, \bibinfo {author} {\bibfnamefont {S.}~\bibnamefont
  {{Dement'ev}}}, \bibinfo {author} {\bibfnamefont {E.}~\bibnamefont
  {{Platacis}}}, \bibinfo {author} {\bibfnamefont {A.}~\bibnamefont
  {{Cifersons}}}, \bibinfo {author} {\bibfnamefont {G.}~\bibnamefont
  {{Gerbeth}}}, \bibinfo {author} {\bibfnamefont {T.}~\bibnamefont
  {{Gundrum}}}, \bibinfo {author} {\bibfnamefont {F.}~\bibnamefont
  {{Stefani}}}, \bibinfo {author} {\bibfnamefont {M.}~\bibnamefont
  {{Christen}}}, \bibinfo {author} {\bibfnamefont {H.}~\bibnamefont
  {{H{\"a}nel}}}, \ and\ \bibinfo {author} {\bibfnamefont {G.}~\bibnamefont
  {{Will}}},\ }\href@noop {} {\bibfield  {journal} {\bibinfo  {journal} {\prl}\
  }\textbf {\bibinfo {volume} {84}},\ \bibinfo {pages} {4365} (\bibinfo {year}
  {2000})}\BibitemShut {NoStop}%
\bibitem [{\citenamefont {{Gailitis}}\ \emph {et~al.}(2004)\citenamefont
  {{Gailitis}}, \citenamefont {{Lielausis}}, \citenamefont {{Platacis}},
  \citenamefont {{Gerbeth}},\ and\ \citenamefont {{Stefani}}}]{gailitis2004}%
  \BibitemOpen
  \bibfield  {author} {\bibinfo {author} {\bibfnamefont {A.}~\bibnamefont
  {{Gailitis}}}, \bibinfo {author} {\bibfnamefont {O.}~\bibnamefont
  {{Lielausis}}}, \bibinfo {author} {\bibfnamefont {E.}~\bibnamefont
  {{Platacis}}}, \bibinfo {author} {\bibfnamefont {G.}~\bibnamefont
  {{Gerbeth}}}, \ and\ \bibinfo {author} {\bibfnamefont {F.}~\bibnamefont
  {{Stefani}}},\ }\href {\doibase 10.1063/1.1666361} {\bibfield  {journal}
  {\bibinfo  {journal} {\pop}\ }\textbf {\bibinfo {volume} {11}},\ \bibinfo
  {pages} {2838} (\bibinfo {year} {2004})}\BibitemShut {NoStop}%
\end{thebibliography}
\end{document}